\documentclass{iau} 
\usepackage{graphicx}

\title[IAUS 317.~~GCs as halo tracers of central cluster galaxies] 
{Globular clusters as tracers of the halo assembly of nearby central cluster 
galaxies}

\author[Michael Hilker \& Tom Richtler]   
{Michael Hilker$^1$
\and Tom Richtler$^2$}

\affiliation{$^1$European Southern Observatory, Karl-Schwarzschild-Str.\,2, \\ 
D-85748, Garching bei M\"unchen, Germany \\ email: {\tt mhilker@eso.org}
\\[\affilskip]
$^2$Departamento de Astronom\'ia, Universidad de Concepci\'on,
Concepci\'on, Chile \\email: {\tt tom@astroudec.cl}}

\pubyear{2015}
\volume{317}  
\jname{The General Assembly of Galaxy Halos: Structure, Origin and Evolution}
\editors{A. Bragaglia, M. Arnaboldi, M. Rejkuba \& D. Romano, eds.}
\begin{document}

\maketitle

\begin{abstract}
The properties of globular cluster systems (GCSs) in the core of the nearby
galaxy clusters Fornax and Hydra\,I are presented. In the Fornax cluster 
we have gathered the largest radial velocity sample of a GCS system so far,
which enables us to identify photometric and kinematic sub-populations
around the central galaxy NGC\,1399. Moreover, ages, metallicities and
[$\alpha$/Fe] abundances of a sub-sample of 60 bright globular clusters
(GCs) with high S/N  spectroscopy show a multi-modal distribution in the
correlation space of these three parameters, confirming heterogeneous
stellar populations in the halo of NGC\,1399.
In the Hydra\,I cluster very blue GCs were identified. They are 
not uniformly distributed around the central galaxies. 3-color photometry
including the $U$-band reveals that some of them are of intermediate age.
Their location coincides with a group of dwarf galaxies under
disruption.  This is evidence of a structurally young stellar halo ``still 
in formation'', which is also supported by kinematic measurements of
the halo light that point to a kinematically disturbed system.
The most massive GCs divide into generally more extended ultra-compact dwarf
galaxies (UCDs) and genuine compact GCs. In both clusters, the spatial
distribution and kinematics of UCDs are different from those of genuine GCs.
Assuming that some UCDs represent nuclei of stripped galaxies, the properties
of those  UCDs can be used to trace the assembly of nucleated dwarf galaxies
into the halos of central cluster galaxies. We show via semi-analytical approaches
within a cosmological simulation that only the most massive UCDs in
Fornax-like clusters can be explained by stripped nuclei, whereas the
majority of lower mass UCDs belong to the star cluster family. 
\end{abstract}

\firstsection 
\section{Introduction}

Central cluster galaxies host systems of thousands of globular clusters (GCs)
which populate their halos out to several tens of effective radii. They are
good probes to trace the assembly history of the diffuse and extended stellar
halos residing in the cores of galaxy clusters. The colors and spectral
line indices of GCs can be used to identify and characterize sub-populations 
of metal-poor and metal-rich as well as young GCs. Together with kinematic
information from large radial velocity samples of GCs one can reconstruct
the assembly history of different halo components. 

The most nearby galaxy clusters, for which their central GC systems (GCSs)
have been photometrically and/or spectroscopically studied in detail, are
Virgo (e.g., \cite[Durrell et al. 2014]{Durrell14}, \cite[Romanowsky et al.
2012]{Romanowsky12}), Fornax (e.g., \cite[Bassino et al. 2006]{Bassino06},
\cite[Schuberth et al. 2010]{Schuberth10}), Hydra\,I (e.g. \cite[Hilker
2002]{Hilker02}, {\cite[Misgeld et al. 2011]{Misgeld11}, \cite[Richtler et al.
2011]{Richtler11}), and Centaurus (e.g., \cite[Hilker 2003]{Hilker03},
\cite[Mieske et al. 2009]{Mieske09}). The number of radial velocity confirmed
GCs around the central galaxies reaches ~1000 GCs for M87 in Virgo and
NGC\,1399 in Fornax. The central GCSs are characterized by bimodal color
distributions. Blue
GCs are commonly interpreted to be metal-poor, and due to their extended
spatial distribution are regarded as good tracers of the old, metal-poor halo
of the central galaxies.  

The high mass end of the GC mass function is dominated by so-called
ultra-compact dwarf galaxies (UCDs), which first have been discovered
in the core of the Fornax cluster (\cite[Minniti et al. 1998]{Minniti98},
\cite[Hilker et al. 1999]{Hilker99}, \cite[Drinkwater et al.
2000]{Drinkwater00}). It is under debate, which
fraction of UCDs originate from stripped nucleated (dwarf) galaxies. 
The kinematic and stellar population properties of stripped nuclei-UCDs
can be used to constrain the contribution of disrupted satellite galaxies
to the build-up of central cluster galaxy halos. The most convincing
case of a stripped nuclei origin of a UCD is the discovery of a supermassive
black hole in M60-UCD1, one of the most massive and densest UCDs in
the Virgo cluster (\cite[Seth et al. 2014]{Seth14}). The supermassive black
hole comprises 15\% of the UCD's total mass.

\begin{figure}[t]
\begin{center}
\includegraphics[width=13.5cm]{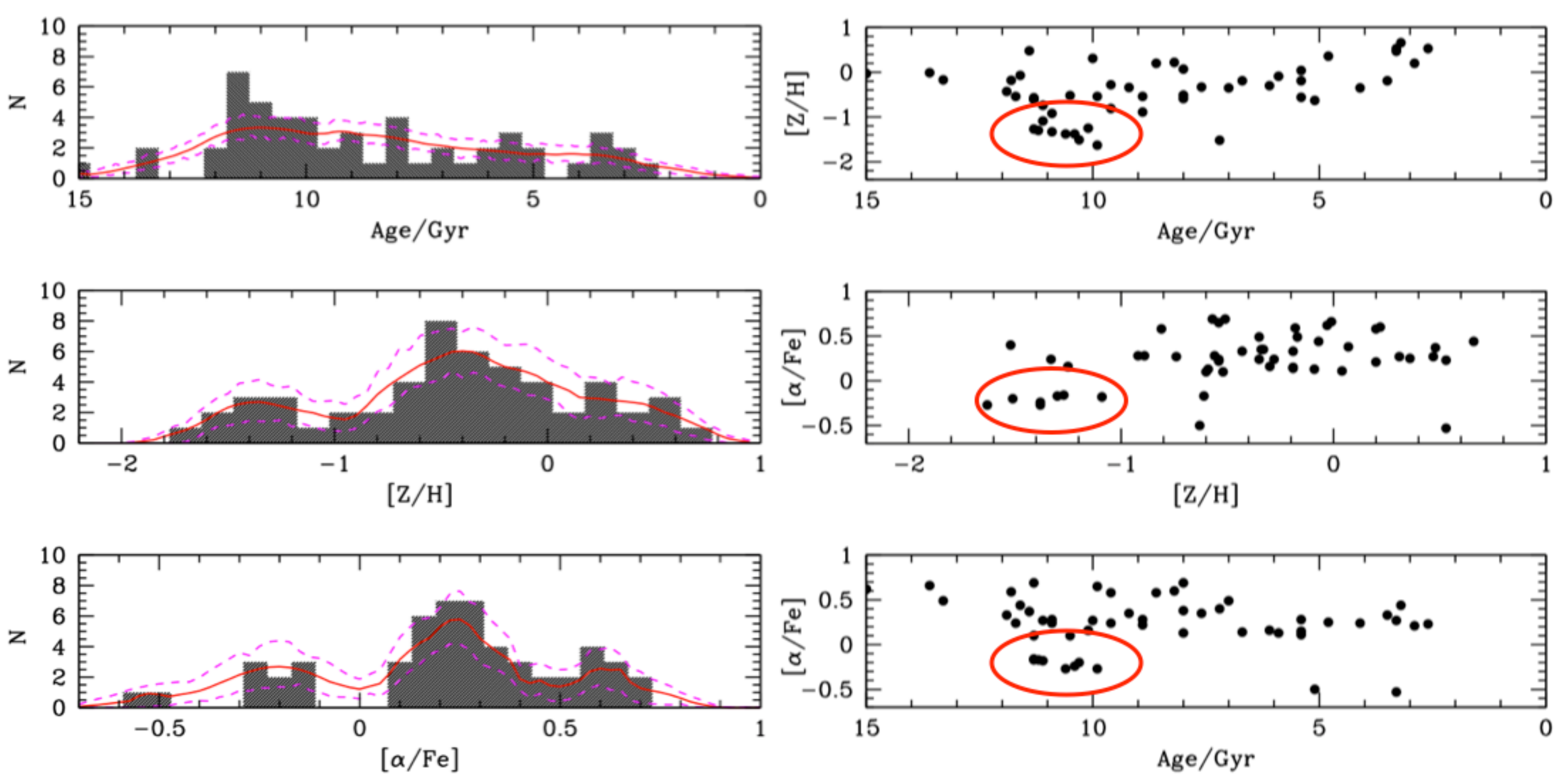} 
\caption{Left panels: distribution of ages, metallicities and [$\alpha$/Fe]
abundances for $\sim50$ Fornax GCs with high S/N spectroscopy.
Right panels: correlations of the three parameters shown in the left.
A distinct 'chemical' sub-group is highlighted by a red ellipse.}
\label{fig1}
\end{center}
\end{figure}

\section{The globular cluster system of the Fornax cluster}

The Fornax cluster has a very well studied GC population. GC counts from
photometric surveys revealed that there exist $\sim6,450\pm700$ GCs
within 83\,kpc projected distance around the central cluster galaxy
NGC\,1399 (\cite[Dirsch et al. 2003]{Dirsch03}). Within 300\,kpc of
NGC\,1399 the GC number counts increase to $\sim11,100\pm2,400$ GCs
(\cite[Gregg et al. 2009]{Gregg09}, derived from the data of \cite[Bassino
et al. 2006]{Bassino06}). The color distribution of GCs in Fornax is bimodal.
The spatial distribution of red (mostly metal-rich) GCs follows the light
of the central galaxy, whereas blue (mostly metal-poor GCs) are more
widely distributed, suggesting that they trace the more extended halo
population of NGC\,1399. 

\cite[Schuberth et al. 2010]{Schuberth10} analysed the kinematics of
$\sim700$ GCs around NGC\,1399 out to 80\,kpc. They found that
bimodality also exists in the kinematic properties of red and blue GCs.
Red GCs are 'well-behaved', showing a gently decreasing velocity
dispersion profile with increasing galactocentric distance. In contrast,
blue GCs have a a generally higher velocity dispersion at all distances
with apparent substructures in the velocity vs. distance diagram.
This might be an imprint of recent galaxy interaction or merging events
that leave accreted (mostly blue) GCs unmixed in phase space. Possible
donor galaxies of accreted halo GCs are the nearby giant early-type
galaxies  NGC\,1404 (see simulations by \cite[Bekki et al. 2003]{Bekki03})
and NGC\,1387, but also dwarf galaxies that got entirely disrupted.

New spectroscopic surveys of GCs in the core region of the Fornax
cluster are being analysed. A large VLT/VIMOS multi-object survey (PI:
Capaccioli) covering the central square degree of the cluster will more
than double the sample of radial velocity members. This will allow us
to search for kinematic substructures in the halo of NGC\,1399. 

Deep VLT/FORS2 spectroscopy on $\sim50$ bright GCs ($M_V<-9$ mag) 
allowed us to measure Lick indices and thus derive their ages,
metallicities and [$\alpha$/Fe] abundances (Hilker, Puzia, et al.,
in preparation). In Fig.\,\ref{fig1} we show the distributions and
correlations of these three quantities. Besides very old GCs there
also exist a sizable number of metal-rich intermediate-age GCs
(2-7 Gyr). And a striking feature in the correlation plots is a distinct
group of seven old metal-poor GCs with sub-solar [$\alpha$/Fe]
abundances. These GCs are restricted in projected galactocentric
distance to NGC\,1399 to a range of 18-31\,kpc, five of them even
to a range of 21-26\,kpc, and thus might represent a 'chemo-dynamical
sub-group' pointing to a common progenitor galaxy. Further kinematic
analysis and correlations in phase space have to show whether this
statement holds true.

\section{Blue and {\it blue} globular clusters in the Hydra\,I cluster}

The central galaxy NGC\,3311 of the Hydra\,I cluster possesses
a very rich GCS ($\sim16,000$ GCs, \cite[Wehner et al. 2008]{Wehner08})
and a large population of UCDs (\cite[Misgeld et al. 2011]{Misgeld11}).
Dynamical analysis of 118 bright GCs and the light around NGC\,3311
revealed a steeply rising velocity dispersion profile, reaching values of
800\,km\,s$^{-1}$ at 100\,kpc galactocentric distance (\cite[Richtler et
al. 2011]{Richtler11}), comparable to the velocity dispersion of the cluster
galaxies. This might either point to a massive dark halo around NGC\,3311
or indicate kinematic substructure in the halo that mimics a dark halo.

Hilker (2002, 2003) noticed that there exist very blue GC candidates around
NGC\,3311, with $0.70<(V-I)=0.85$ mag, much bluer than the blue peak of
metal-poor GCs ($(V-I)_{\rm blue}=0.9$). Those GCs are not centred on
the galaxy but show an offset towards the north-east. Their blue $(V-I)$
color can be interpreted in two ways. Either these GCs are old but very
metal-poor, with [Fe/H]$\simeq-2.5$ dex, or they are metal-rich but rather
young (1-5\,Gyr). In order to break the age-metallicity degeneracy in the
$(V-I)$ color, and thus uncover the nature of those blue GCs, we took
$U$-band images in the cluster core with FORS at the VLT. In Fig.\,\ref{fig2}
we show the distribution of GCs in the $(U-V)$-$(V-I)$ 2-color diagram
(left panels). With help of the PARSEC model grid for single stellar populations
(\cite[Bressan et al. 2012]{Bressan12}) one can select GCs of different ages
and/or metallicities and study their spatial distributions (right panels in
Fig.\,\ref{fig2}). Whereas old GCs are mostly centred on NGC\,3311, the
distribution of young GCs ($<2$\,Gyr) is totally different. Most of them 
are displaced towards the North, coinciding with the location of a group
of dwarf galaxies (\cite[Misgeld et al. 2008]{Misgeld08}). Others are
located south-east of NGC\,3311 in the wake of the spiral galaxy
NGC\,3312, which is cruising through the Hydra\,I cluster at high speed,
as evidenced by its compressed eastern edge.

The offset distribution of young, blue GCs coincides with the location of
an offset faint stellar envelope around NGC\,3311 (\cite[Arnaboldi et al. 
2012]{Arnaboldi12}), an offset X-ray halo (\cite[Hayakawa et al.
2004]{Hayakawa04}) and a region of high velocity dispersion in the halo
light (Hilker et al., in prep.). Taken all together, this points to a
non-equlibrium situation in the core of Hydra\,I. The central galaxy
seems not to be at rest with the cluster potential well, and infalling
substructure is building up the central halo and intra-cluster light.
Thus we are witnessing 'ongoing formation' of a central cluster halo.

\begin{figure}[h!]
\begin{center}
\includegraphics[width=13.5cm]{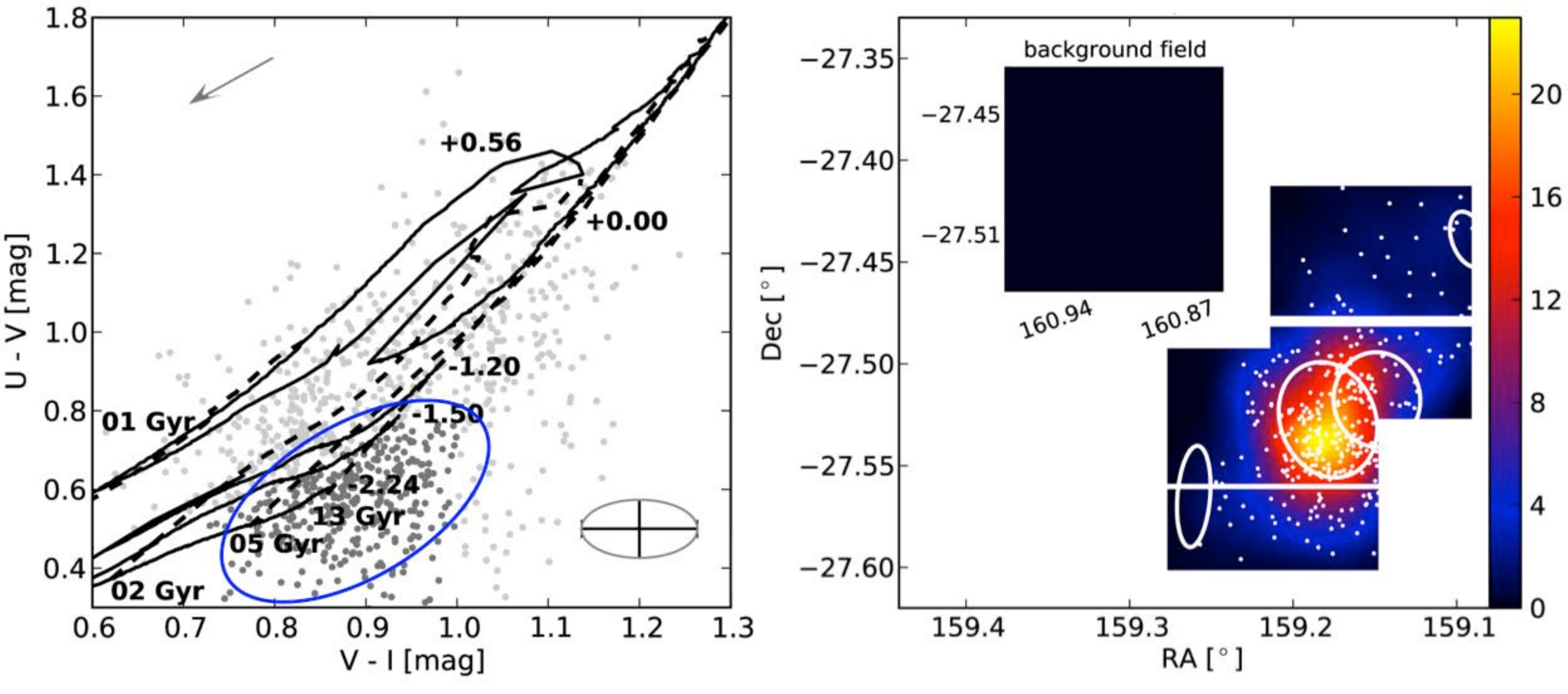} 
\includegraphics[width=13.5cm]{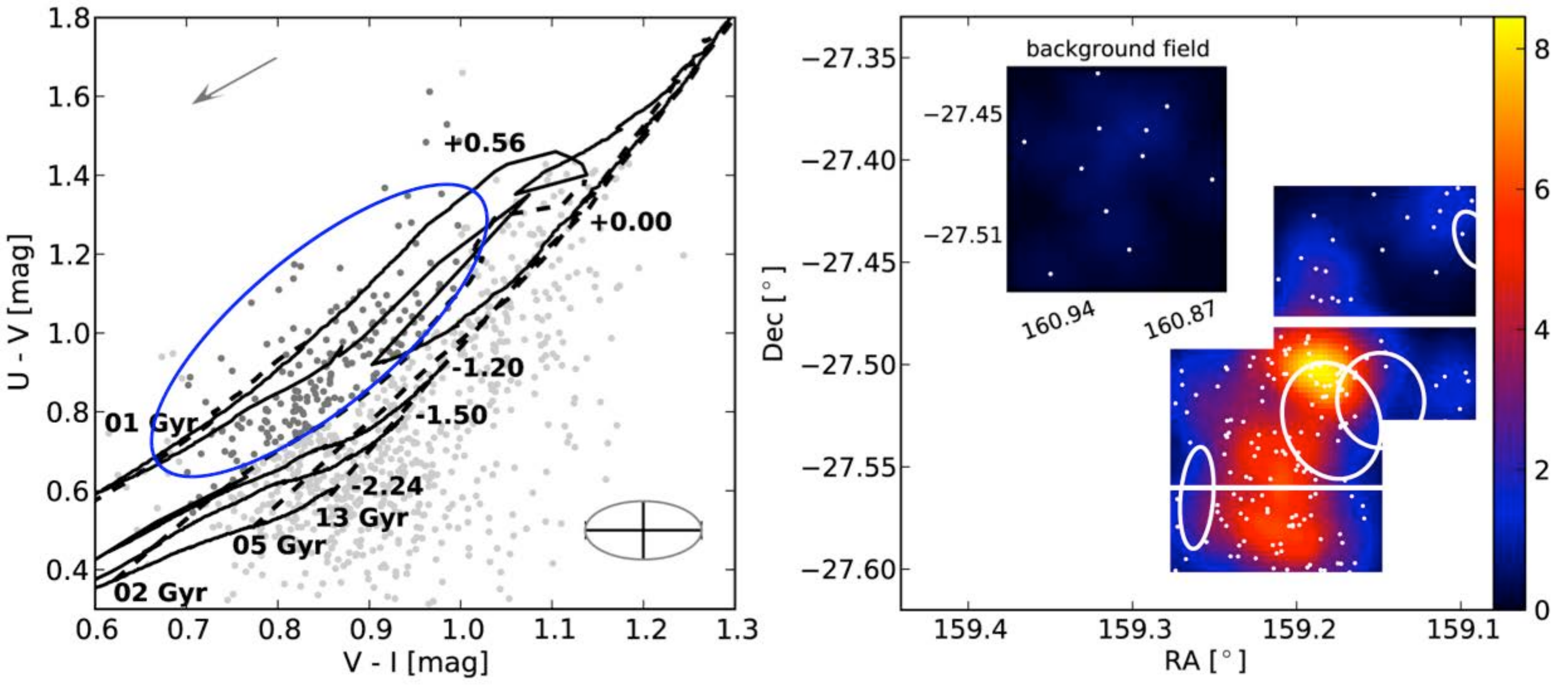} 
\caption{{\it Left panels}: GCs around NGC\,3311 in Hydra\,I in the 
$(U-V)$-$(V-I)$ color-color space.  Single stellar population tracks
from PARSEC models (\cite[Bressan et al. 2012]{Bressan12}) for
various ages (dashed lines) and metallicities (solid lines) are overlaid.
In the upper panel, the region of old GCs (dark grey dots) is highlighted
by a blue ellipse, in the lower panel the region for young ($<$2\,Gyr) GCs.
{\it Right panels}: spatial distribution of the selected old, metal-poor
GCs (upper panel) and the young GCs (lower panel). White ellipses
indicate the central major galaxies in Hydra\,I. The scale on the
right are numbers per square arcmin. Note that the comparison
background field is located several degrees east of the cluster.}
\label{fig2}
\end{center}
\end{figure}

\section{The most massive GCs=UCDs: two formation channels}

As mentioned in the introduction, very massive GCs cannot
easily be distinguished from so-called ultra-compact dwarf galaxies (UCDs).
One should rather think of the different formation channels that can
lead to rather compact ($r_{\rm eff}=3$-100\,pc) objects in the mass
range $10^6<M<10^8 M_{\odot}$. One viable channel is the disruption
of nucleated (dwarf) galaxies on radial orbits that pass the central cluster
galaxies at small perigalactic radii and leave a 'naked' UCD-like stripped
nucleus behind. However, cosmological simulations combined with a
semi-analytic description to identify disrupted satellite galaxies suggest
that this only explains the observed number of UCDs more massive than
$M>10^{7.3} M_{\odot}$ (\cite[Pfeffer et al. 2014]{Pfeffer14}). The observed
number of lower mass UCDs is much larger than that of predicted stripped
nuclei. They should, thus, be of star cluster origin, either formed as very
massive genuine globular cluster (e.g., \cite[Murray 2009]{Murray09}) or
being the result of merged star cluster complexes (\cite[Fellhauer \& Kroupa
2002]{Fellhauer02}).

In order to constrain the formation of UCDs we have studied the structural
composition and clustering properties of 97 UCDs in the halo of NGC\,1399,
the central Fornax galaxy (see contribution by Voggel, Hilker \& Richtler,
this volume). We found evidence for faint stellar envelopes around several
UCDs with effective radii of up 90\,pc. One particularly extended UCD
shows clear signs of tidal tails extending out to $\sim$350\,pc (see
Fig.\,\ref{fig3}, left panel). This is the first time that a tidal tail has been
detected around a UCD.
But the most striking result is that we detect, in a statistical sense, a
local overdensity of GCs on scales of $\leq1$\,kpc around UCDs.
In particular blue (likely metal-poor) GCs are clustered around UCDs.
These could either be remnant GCs of a formerly rich GCS around a 
disrupted nucleated dwarf galaxy, or surviving star clusters of a merged
super star cluster complex (e.g., \cite[Br\"uns et al. 2009]{Bruens09}).
A remote UCD, 85\,kpc south of NGC\,1399, possesses four GC candidates
within 1 kpc radius, well within its tidal radius of 1.36\,kpc, but shows
no signs of a faint envelope in the same radius (see Fig.\,\ref{fig3}).
The nature of this configuration is intriguing, pointing to a progenitor
object that had a very rich substructure, maybe a complex star cluster
system formed in- or outside a former host galaxy. Radial velocity
measurements have to show whether the companion GCs are indeed
bound to the host UCD.

\begin{figure}[t]
\begin{center}
\includegraphics[width=12.2cm]{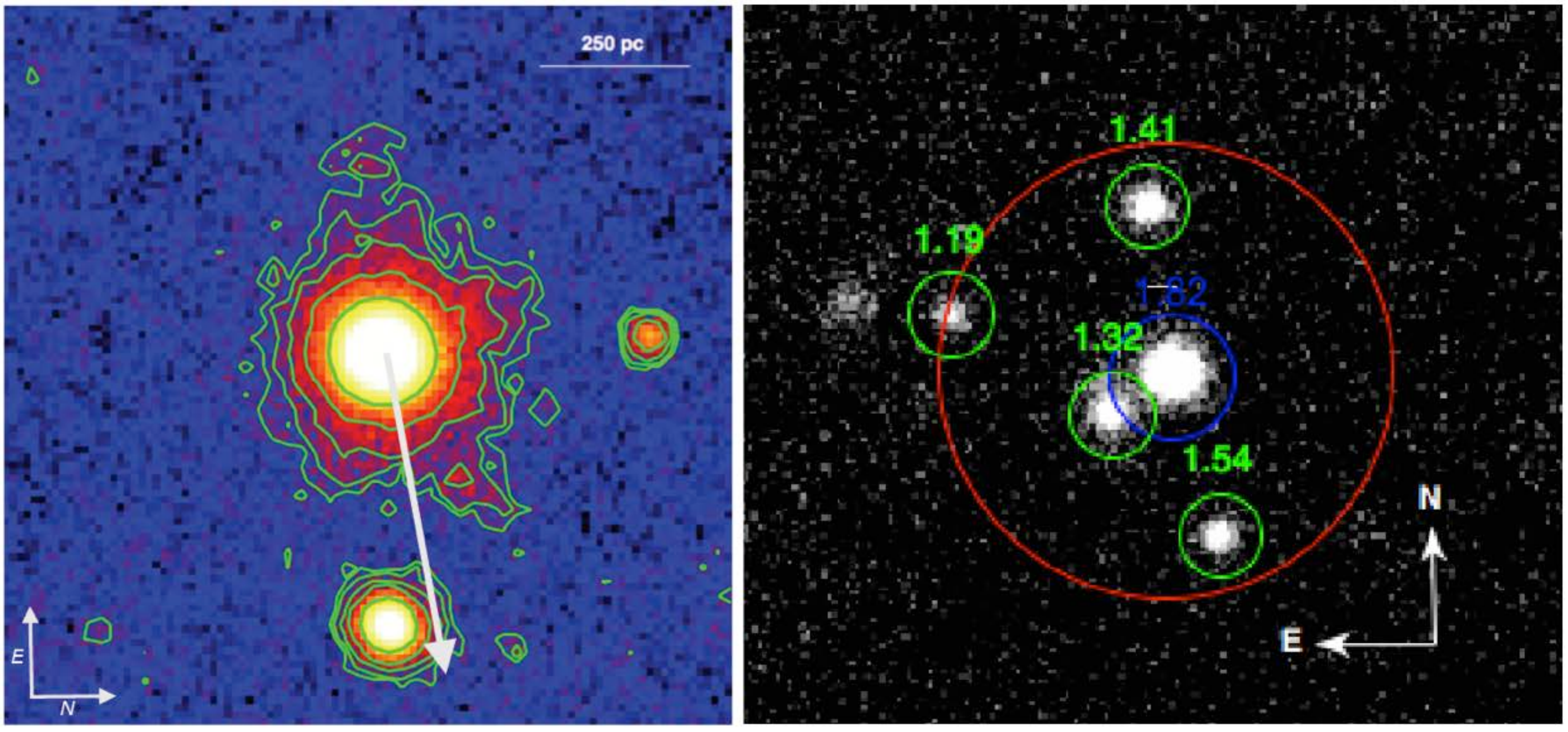} 
\caption{Left: UCD in the Fornax cluster that shows prominent tidal
tail-like structures that extend out to 350\,pc. The arrow indicates the
direction towards the central Fornax galaxy NGC\,1399. 
Right: Remote UCD (blue circle) at 85\,kpc distance to NGC\,1399 that
harbours four GC candidates (green circles) within 1\,kpc (red circle).
The objects are labeled with their Washington $C-T_1$ colors (from 
\cite[Dirsch et al. 2003]{Dirsch03}). The figures are taken from Voggel,
Hilker \& Richtler (2015, submitted).}
\label{fig3}
\end{center}
\end{figure}

\section{Summary and outlook}

Our general conclusions from the findings in the Fornax and Hydra\,I
galaxy clusters shown in this contribution can be summarized as follows:

\begin{itemize}
\item In general, old globular clusters are good tracers of spheroid
(red GCs) and halo (blue GCs) populations of ellipticals.
\item The predominant GC population in the outer halo regions
are the blue GCs. They trace the halo assembly history.
\item Kinematics together with metal abundances of GCs is a powerful 
tool to find substructures and trace recent accretion events.
\item In an appropriate 3-color space, sub-populations of blue GCs
can be identified. Blue is not the same as {\it blue}!
\item Ultra-compact dwarf galaxies (=the most massive star clusters) 
are a mixed bag of objects: most of them ($>$80\%) are of star
cluster origin.
\item Extended stellar envelopes and overdensities of star clusters 
around them might hint to the accretion of nucleated dwarf galaxies
or evolved super star cluster complexes.
\end{itemize}

In the coming years we will see great progress in extra-galactic globular
cluster science. The Virgo and Fornax clusters are being scrutinized by
photometric multi-wavelength wide-field surveys, covering the $U$- to
the $K$-band. GCs and UCDs serve as one of the main tracers of the
spatial distribution of baryonic structure in these clusters. Massive
spectroscopic follow-up surveys are or will be launched to collect radial
velocities and element abundances of thousands of GCs and UCDs
around the central cluster galaxies in order to find chemo-dynamical
substructures that constrain their halo assembly history.

\end{document}